# Thermal Radiation at the Nanoscale and Applications

Pierre-Olivier Chapuis[1], Bong Jae Lee[2], and Alejandro Rodriguez[3]

[1] Univ Lyon, CNRS, INSA-Lyon, Université Claude Bernard Lyon 1, CETHIL UMR5008, F-69621, Villeurbanne, France
E-mail : olivier.chapuis@insa-lyon.fr

[2] Department of Mechanical Engineering, Korea Advanced Institute of Science and Technology, Daejeon 34141, South Korea

[3] Department of Electrical Engineering, Princeton University, Princeton, New Jersey 08544, USA

Abstract=====================================================================

There has been a paradigm shift from the well-known laws of thermal radiation derived over a century ago, valid only when the length scales involved are much larger than the thermal wavelength (around 10 µm at room temperature), to a general framework known as fluctuational electrodynamics that allows calculations of radiative heat transfer for arbitrary sizes and length scales. Near-field radiative heat transfer and thermal emission in systems of sub-wavelength size can exhibit super-Planckian behaviour, i.e. flux rates several orders of magnitude larger than that predicted by the Stefan–Boltzmann (or blackbody) limit. These effects can be combined with novel materials, e.g. low-dimensional or topological systems, to yield even larger modifications and spectral and/or directional selectivity. We introduce briefly the context and the main steps that have led to the current boom of ideas and applications. We then discuss the original and impactful works gathered in the associated Special Topic collection, which provides an overview of the flourishing field of nanoscale thermal radiation.

==========================================================================

The ability to control thermal radiation is important for a broad range of applications, including thermal management, spectroscopy, optoelectronics, and energy-conversion devices. Many of these applications can take advantage of nanotechnology, often by nano-manufacturing certain components which are involved in the technologies or making them more compact. As a result, there is a strong need for an accurate description of thermal radiation in nanoscale configurations. Furthermore, the general approach to investigate these phenomena has undergone a paradigm shift, from the well-known laws of thermal radiation, valid only when the involved length scales are much larger than the thermal wavelength (around 10 µm at room temperature), to a general framework known as fluctuational electrodynamics that allows calculations of radiative heat transfer for arbitrary sizes and distances. In the following, we first describe some of the remarkable steps that led to the current state of the art in thermal radiation engineering, then provide key concepts explored by the diverse works reported in the Special Topic collection entitled 'Thermal Radiation at Nanoscale and Applications', which highlights the dramatic surge in both theoretical and applied investigations of this field.

==

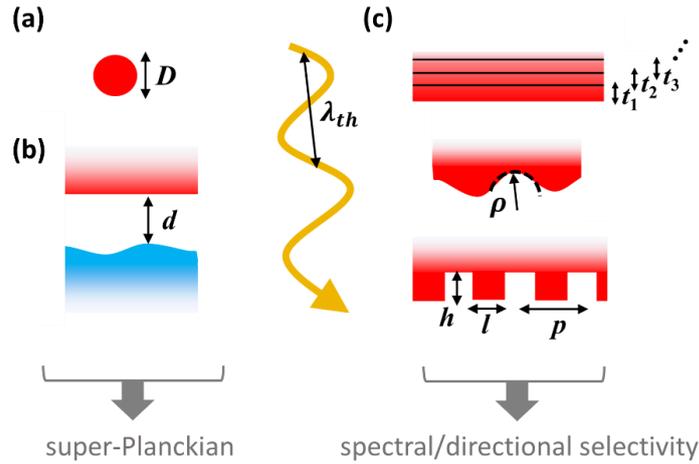

**Figure 1.** *Examples of size effect in thermal radiation (a-c). The thermal wavelength $\lambda_{th}$ is schematized in yellow. (a) Thermal emission by an object of sub-wavelength size $D \ll \lambda_{th}$. (b) Near-field thermal radiation, also called thermal-photon tunneling, where $d \ll \lambda_{th}$ is the distance between the radiating objects at different temperatures. (c) Three examples of large surfaces emitting thermal radiation downwards with features of size comparable to or smaller than $\lambda_{th}$: a multilayer with layers of thicknesses $t_i$, a surface with a roundy shape of curvature radius $\rho$, a metasurface or a grating with height $h$, pillar length $l$ and periodicity $p$*

Planck's famous law of surface-to-surface radiative exchange between opaque bodies is a century old[1] and is able to deal successfully with innumerable configurations. However, Planck himself underlined in his book that the law would only be able to address objects, distances and curvature radii larger than the relevant wavelengths at which radiative transfer occurs. As a consequence, other famous features of macroscopic thermal radiation, such as the $T^4$ dependence of Stefan-Boltzmann's law, are not expected to work at small scales. Fifty years ago, two landmark papers went further and correctly described radiative heat transfer in situations outside the aforementioned ray optics regimes, namely an object of subwavelength size[2] (1970, see Fig. 1a) and two objects separated by a small vacuum gap[3] (1971, see Fig. 1b). To do so, they relied on fluctuational electrodynamics (FE), a theory combining Maxwell's equations and statistical principles developed by Sergei Rytov[4,5] and coworkers, an approach sometimes referred to as stochastic electrodynamics.

Because FE exploits the full generality of Maxwell's equations in describing thermal radiation, wave effects such as interference and photon tunneling are included in the theory. The electromagnetic waves carrying thermal radiation originate from classical albeit stochastic sources describing thermal agitation of charges in matter. A key element of this formulation is the fluctuation-dissipation theorem (FDT), expressed by Callen and Welton[6] following the description of random fluctuation of charges in conductors, i.e. the electric noise, by Nyquist[7] and Johnson[8]. The FDT provides a link between dissipation (how energy is absorbed in a medium) and thermal agitation, underlining the fact that energy is a quadratic quantity described by two-point correlation functions. Green's functions relating causes - thermal sources - and consequences - electromagnetic fields - provide an additional element allowing studies of arbitrary structural configurations. In addition to radiative heat transfer, FE is also at the heart of the nanoscale description of other phenomena, such as the Casimir force[9] and noncontact friction[10].

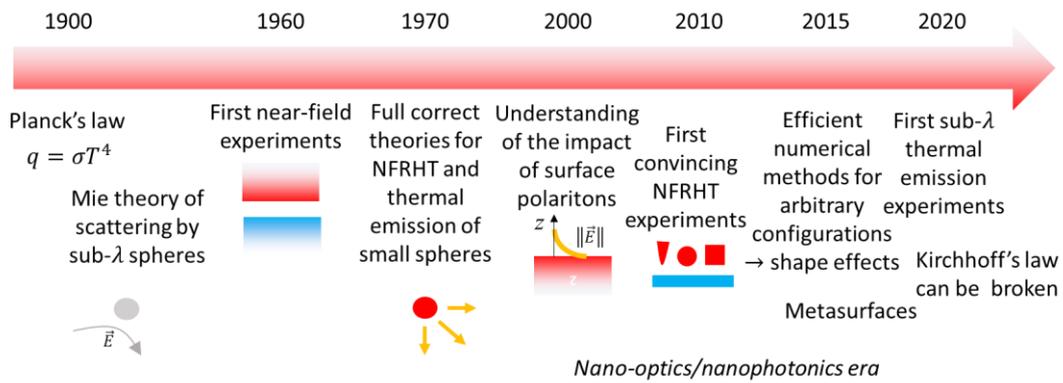

**Figure 2**. *Timeline of advances in the field of thermal radiation at nanoscale*

While the theoretical principles underlying radiative heat transfer (RHT) were established 50 years ago, experiments were not easy to realize at the time due to the small spatial scales required: at room temperature, the thermal wavelength is close to 10 µm, and small-scale effects become relevant only in the sub-micrometre regime. Early attempts to measure near-field radiative heat transfer (NFRHT) for NASA, performed at lower temperatures (recall Wien's law where the peak thermal wavelength is $\lambda_{th} \approx 3000/T$ µm, with $T$ the temperature) and therefore much longer wavelengths[11], or at Philipps Research, where the experiment by Hargreaves[12] (supervised by Hendrik Casimir, the colleague of Dirk Polder) predated the landmark theoretical paper, provided first hints but could not lead to numerous experimental confirmations. It is only in the last 15 years (see initial papers by Shen *et al.* and Rousseau *et al.* in 2008[13,14]) that a profusion of sensitive near-field experiments appeared[15], as a consequence of the development of nanotechnology with atomic force microscopy, nanolithography and MEMS fabrication processes. The first clear experiments of thermal emission of sub-wavelength objects are very recent, less than 5-year old[16,17]. In both cases, it was shown that the radiated flux can exceed that predicted by Planck's blackbody theory (applicable only in the ray optics regime mentioned above, but unfortunately applied often out of its validity domain). Such a phenomenon has been termed super-Planckian emission[18].

Just prior, a theoretical revival emerged 20 years ago, when it was realized that surface polaritons, i.e. collective charge oscillations at surfaces associated with bound material resonances, could introduce interesting features. One of them is associated with coherent thermal radiation: scattering surface polaritons by a periodic structure (a grating) allows for directional emission at each contributing wavelength[19]. This is in contrast to the usual broadband and isotropic nature of far-field emission. Surface nanostructuring in optics has led to the field of metasurfaces[20], proving a fruitful an avenue for novel thermal-emission engineering. As an example, spectrally and/or directionally selective emission have become possible. More strikingly, it was shown[21] recently that bi-anisotropic materials can break, under certain conditions, the famous Kirchhoff's law[22], a pedestal of thermal radiation studies, which states that spectral-directional emissivity is equal to spectral-directional absorptivity[23]. Fig. 2 summarizes graphically some of the key dates mentioned above associated with the field of nanoscale thermal radiation.

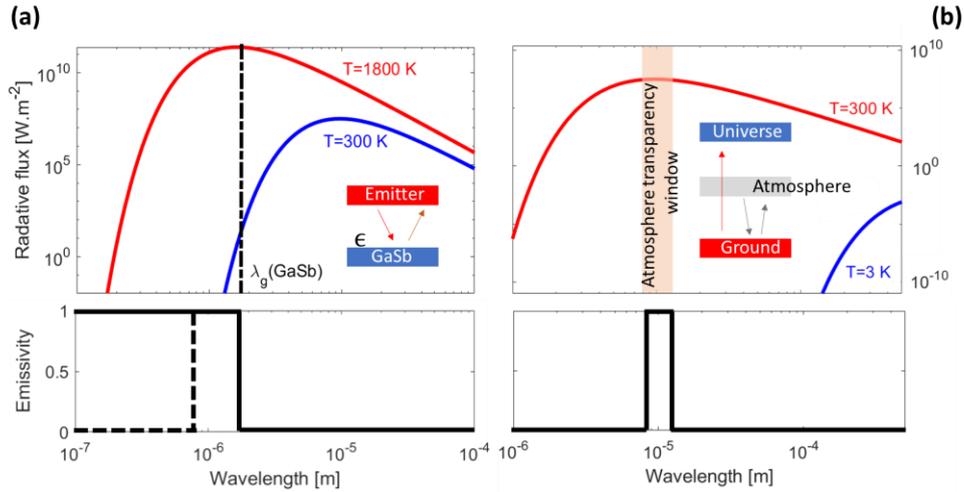

**Figure 3**. *Spectral selectivity required for two key applications: (a) thermophotovoltaics, here with a GaSb cell at room temperature and an emitter at 1800 K, (b) night-time radiative cooling. Blackbodies at the different temperatures are represented in red and blue. In TPV (a), high efficiency can be achieved if the cell emissivity is unity close to the bandgap since reflected photons are not lost. Reducing the spectral bandwidth however decreases the output power density. Radiative cooling (b) takes place if the body radiates toward universe (low temperature) while reflecting other radiative fluxes. For day-time radiative cooling, solar radiation (not represented here) should especially be reflected.*

In parallel to all these fundamental developments, there have also been several forays into thermal applications. When in the near field, surface polaritons lead to spectra very different from those usually known in the far field. Close-to-monochromatic spectra can be obtained for small distances or small emitters[24,25]. It was postulated early[26,27] that this could be helpful for thermophotovoltaics (TPV), one among other compelling applications of NFRHT. TPV in the far field (see Fig. 3(left)) involves conversion of thermal radiation from a hot emitter into electricity – photovoltaics operating in the infrared. One hurdle of solar photovoltaics is the need to convert a broad radiative spectrum while photovoltaic cells work efficiently for radiation of energy confined just above the bandgap. In contrast, the TPV efficiency is better controlled as the incoming radiation can be confined spectrally. Despite it being a very mature field (Kolm and Aigrain are credited with the first steps in the 1950s-1960s), TPV[28] design is currently experiencing significant interest owing to advances in nanofabrication[29,30], development of back-reflectors allowing for high efficiency by recycling nonconverted photons[31–33] and the first experimental demonstrations of near-field TPV conversion in the last 5 years[34–37]. While near-field TPV experiments have thus far failed to exploit surface-polariton effects, several ongoing efforts show promise. The energy crisis highlights indeed the need for recovering waste heat at all temperature scales.

Another key application benefitting from theoretical progresses in tailoring far-field thermal radiative properties is radiative cooling[38–41] (see Fig. 3(right)). Radiative cooling consists in emitting more thermal radiation than absorbing it, and therefore usually requires a strong emission in the atmospheric transmission window in the mid-infrared band. In some sense, it is the opposite of the greenhouse effect. While this is quite an old topic, the possibility of nanostructuring thermal emitters has broadened the panel of concepts that can be applied for enhancing the effect in day-time environments[42]. It is especially timely due to the need for passive cooling of buildings and humans in hot environments.

Finally, all these advances would not have been possible without improvements in metrology, spectroscopy, and nanofabrication. For spectral analysis, this includes progress in near-field spectroscopic techniques based on atomic force microscopy[43–47] combined with the more common Fourier-transform infrared (FTIR) spectrometer, and the possibility of infrared ellipsometry. At the integrated level (power), the development of tiny thermocouples or resistive thermometers has allowed for measurement of sensitive heat flux densities.

==

At this stage, we would like to emphasize that there are many insightful references dealing with thermal radiation at the nanoscale. We wish first to highlight the book by Zhang[48], which provides a detailed introduction. Among good review papers on particular sub-topics, we can mention the following. Small-object emission has been discussed by Carlos-Cuevas *et al*.[49]. Near-field radiative heat transfer was discussed e.g. in Refs [50–52] and more recently by Papadakis *et al*.[53] with a focus on resonances in dielectrics. A report on current experiments can be found in Ref. [15], with Song *et al.* providing a detailed review on near-field thermophotovoltaic energy conversion[54]. Thermal emission of surfaces and metasurfaces was reported in Refs [55,56]. The possibility of designing thermal logics and functions was underlined by Biehs and Ben-Abdallah[57]. Many-body systems, as electromagnetism is nonadditive, are now addressed in the near field[58]. The combination of radiative heat transfer and junctions in energy-conversion devices was detailed by Tervo *et al*.[59]. Many other references could be added.

==

We now turn to an analysis of the topics addressed in the Special Topic collection, which provides a nice overview of the current lines of investigations in the field. Figure 4 summarizes the key contributions, splitting between configurations, methods and applicative fields. One a clear trend toward increasingly complex configurations, which now either couple thermal radiation studies with electron-hole transport in materials or address advanced topologies such as metasurfaces, nanoparticle chains, or higher-dimensiona objects where orientation plays a key role. Related to computational methods, we observe progressively that one-dimensional FE is replaced by numerical calculations and even large-scale brute-force optimization. On the experimental side, nanofabrication is spread among all studies, where spectroscopy is required when spectral selectivity is key to the goal and flux measurements can be realized by photoacoustic or photothermal techniques. Finally, we can divide the applications studied into three categories: (i) purely thermal, such as those involving thermal management (including switching/rectification) or radiative cooling, (ii) those where electrical control or output is desired in a device (bolometers, MOS transistors, PIN diodes, energy-harvesting…), and (iii) those where coupling between thermal radiation and other fluctuating phenomena (such as near-field friction) is considered. We note that the articles of the collection are published under many categories of APL – metasurfaces/materials[60–67], photonics/optoelectronics[68–73], properties[74–78], energy[79–81], device physics[82–84], imaging[85,86], applied physics[87], surfaces/interfaces[88] -, which highlights the interdisciplinarity and the various fields addressed by nanoscale thermal radiation.

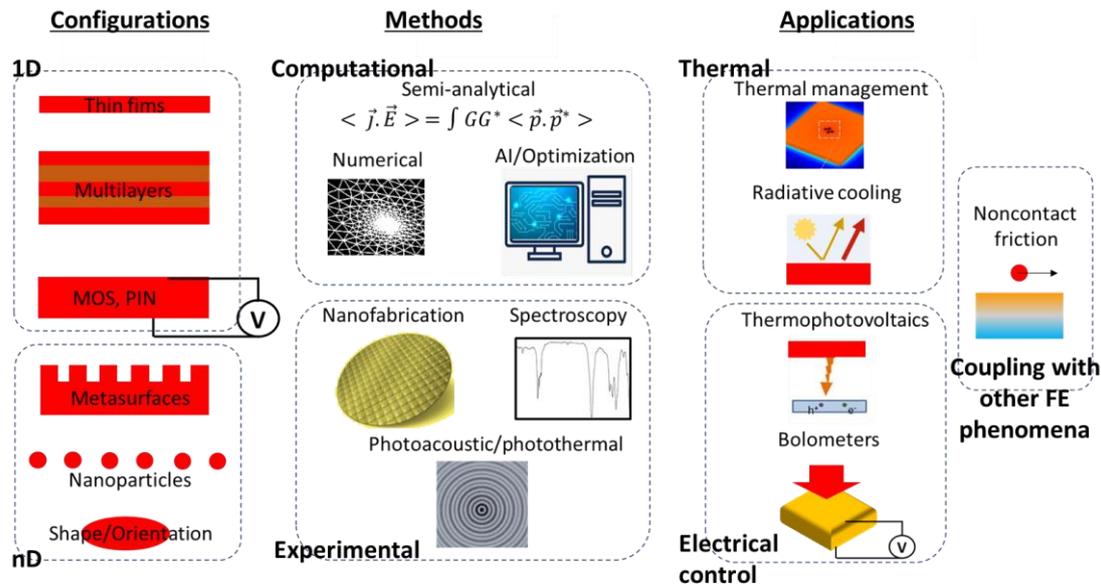

**Figure 4.** *Schematic showing typical configurations addressed, the methods applied, and the applications involved.*

Finally, this Special Topic collection reveals the diversity of specialization area and scientific origin of its contributors. In contrast to early days of the field, more than half of the submissions are coming from Asia, including roughly one third from China. America (USA, Canada, Mexico) is responsible for ¼ of the submissions, while the rest are coming from Europe (France, Germany, Finland, Spain, etc.). While this distribution of the submissions provides probably only a qualitative idea of the forces at the global scale, it is in line with the notable rapid rise of China in optics and condensed matter-related fields and the current strength of other Asian countries (Japan, South Korea). It will be interesting to analyze where applications develop.

Conclusions and prospects

To conclude, we underline that the abovementioned sub-topics highlight very well the dynamism of the community tackling thermal radiation at the crossroad of heat transfer and nanophotonics, as well as the variety of applications that can be addressed. This resonates particularly in this time where the need for rational and optimal use of energy and the quest for efficient harvesting are extremely important. One difficulty we have not discussed yet is the cost and upscaling of the envisioned structures to the level of technological devices. Nanostructuring is not always easy for large-scale elements, and strategies based on bottom-up system design or chemical synthesis would certainly be preferred. Radiative-cooling textiles and paintings have already entered this stage. For thermophotovoltaics, start-ups have already begun to address the question of economic viability. Other application-driven systems have hardly tackled such issues yet. At the level of fundamental science, there remain many questions to be addressed. Many pillars of the macroscopic thermal-radiation theory have been progressively revisited over the past decades: the blackbody limit, Kirchhoff's law, and even nonlinear fluctuation statistics. There certainly remain others soon to undergo their 'revolution'.

Acknowledgements

We thank Profs. L. Cohen and A. Tredicucci for suggesting us to participate in the guest editorship of a Special Topic collection, and Mr J.-I. Rodriguez and the APL staff for the smooth organization. We also thank the APL editors who fully managed the review process for the Special Topic collection.

P.O.C. acknowledges funding from EU projects TPX-Power (2020-EIC-FETPROACT-2019/GA951976) and OPTAGON (H2020-FETOpen-2018-2019-01/GA964698), and ANR projects STORE (ANR-21-CE42-0032) and CASTEX (ANR-21-CE30-0027). B.J.L acknowledges funding supported by the Basic Science Research Program (NRF-2019R1A2C2003605) through the National Research Foundation of Korea (NRF) funded by Ministry of Science and ICT. A.W.R acknowledges funding supported by the Cornell Center for Materials Research (MRSEC) through award DMR-1719875 and the Defense Advanced Research Projects Agency (DARPA) under agreements HR00112090011, HR00111820046, and HR0011047197.

REFERENCES

1. Planck, M., *The Theory of Heat Radiation*. Dover (New-York), 1914.

2. Kattawar, G. W. & Eisner, M. Radiation from a Homogeneous Isothermal Sphere. *Appl Opt* **9**, 2685 (1970).

3. Polder, D. & Van Hove, M. Theory of radiative heat transfer between closely spaced bodies. *Phys Rev B* **4**, 3303–3314 (1971).

4. Rytov, S. M. *Theory of Electric Fluctuations and Thermal Radiation*. (Electronics Research Directorate, Air Force Cambridge Research Center, Air Research and Development Command, U.S. Air Force, 1959).

5. Rytov, S. M., Kravtsov, Y. A. & Tatarskii, V. I. *Principles of Statistical Radiophysics 3 - Elements of random fields*. Springer-Verlag (Berlin Heidelberg), 1989.

6. Callen, H. B. & Welton, T. A. Irreversibility and Generalized Noise. *Physical Review* **83**, 34–40 (1951).

7. Nyquist, H. Thermal Agitation of Electric Charge in Conductors. *Physical Review* **32**, 110–113 (1928).

8. Johnson, J. B. Thermal Agitation of Electricity in Conductors. *Physical Review* **32**, 97–109 (1928).

9. Parsegian, V. A. *Van der Waals Forces: a Handbook for Biologists, Chemists, Engineers, and Physicists.* (Cambridge University Press, 2005).

10. Volokitin, A. I. & Persson, B. N. J. Near-field radiative heat transfer and noncontact friction. *Rev Mod Phys* **79**, 1291–1329 (2007).

11. Domoto, G. A., Boehm, R. F. & Tien, C. L. Experimental Investigation of Radiative Transfer Between Metallic Surfaces at Cryogenic Temperatures. *J Heat Transfer* **92**, 412 (1970).

12. Hargreaves, C. M. Anomalous radiative transfer between closely-spaced bodies. *Phys Lett A* **30**, 491–492 (1969).

13. Narayanaswamy, A., Shen, S. & Chen, G. Near-field radiative heat transfer between a sphere and a substrate. *Phys Rev B Condens Matter Mater Phys* **78**, 115303 (2008).

14. Rousseau, E., Siria, A., Jourdan, G., et al. Radiative heat transfer at the nanoscale. *Nat Photonics* **3**, 514–517 (2009).

15. Lucchesi, C., Vaillon, R. & Chapuis, P.-O. Radiative heat transfer at nanoscale: experimental trends and challenges. *Nanoscale Horiz* **6**, 201 (2021).

16. Shin, S., Elzouka, M., Prasher, R. & Chen, R. Far-field coherent thermal emission from polaritonic resonance in individual anisotropic nanoribbons. *Nat Commun* **10**, 1377 (2019).

17. Thompson, D., Zhu, L., Mittapally, R., et al. Hundred-fold enhancement in far-field radiative heat transfer over the blackbody limit. *Nature* **561**, 216 (2018).

18. Guo, Y., Cortes, C. L., Molesky, S. & Jacob, Z. Broadband super-Planckian thermal emission from hyperbolic metamaterials. *Appl Phys Lett* **101**, 131106 (2012).

19. Greffet, J.-J., Carminati, R., Joulain, K., et al. Coherent emission of light by thermal sources. *Nature* **416**, 61–64 (2002).

20. Yu, N. & Capasso, F. Flat optics with designer metasurfaces. *Nat Mater* **13**, 139–150 (2014).

21. Shayegan, K. J., Biswas, S., Zhao, B., Fan, S. & Atwater, H. A. Direct observation of the violation of Kirchhoff's law of thermal radiation. *Nat Photonics* **17**, 891–896 (2023).

22. Kirchhoff, G. Ueber das Verhältniss zwischen dem Emissionsvermögen und dem Absorptionsvermögen der Körper für Wärme und Licht. *Ann Phys* **185**, 275–301 (1860).

23. Greffet, J.-J. & Nieto-Vesperinas, M. Field theory for generalized bidirectional reflectivity: derivation of Helmholtz's reciprocity principle and Kirchhoff's law. *Journal of the Optical Society of America A* **15**, 2735–2744 (1998).

24. Mulet, J.-P., Joulain, K., Carminati, R. & Greffet, J.-J. Enhanced radiative heat transfer at nanometric distances. *Microscale Thermophysical Engineering* **6**, 209–222 (2002).

25. Narayanaswamy, A. & Chen, G. Surface modes for near field thermophotovoltaics. *Appl Phys Lett* **82**, 3544–3546 (2003).

26. Whale, M. D. & Cravalho, E. G. Modeling and Performance of Microscale Thermophotovoltaic Energy Conversion Devices. *IEEE Transactions on Energy Conversion* **17**, 130–142 (2002).

27. DiMatteo, R. S., Greiff, P., Finberg, S. L., et al. Enhanced photogeneration of carriers in a semiconductor via coupling across a nonisothermal nanoscale vacuum gap. *Appl Phys Lett* **79**, 1894–1896 (2001).

28. Burger, T., Sempere, C., Roy-Layinde, B. & Lenert, A. Present Efficiencies and Future Opportunities in Thermophotovoltaics. *Joule* **4**, 1660–1680 (2020).

29. Wang, Z., Kortge, D., He, Z., *et al.* Selective emitter materials and designs for high-temperature thermophotovoltaic applications. *Solar Energy Materials and Solar Cells* **238**, 111554 (2022).

30. Lenert, A., Bierman, D. M., Nam, Y., *et al.* A nanophotonic solar thermophotovoltaic device. *Nat Nanotechnol* **9**, 126–130 (2014).

31. Omair, Z., Scranton, G., Pazos-Outón, L. M., *et al.* Ultraefficient thermophotovoltaic power conversion by band-edge spectral filtering. *Proc Natl Acad Sci U S A* **116**, 15356–15361 (2019).

32. Fan, D., McSherry, S., Lee, B., *et al.* Near-perfect photon utilization in an air-bridge thermophotovoltaic cell. *Nature* **586**, 237–241 (2020).

33. LaPotin, A., Schulte, K. L., Steiner, M. L., *et al.* Thermophotovoltaic efficiency of 40%. *Nature* **604**, 287–291 (2022).

34. Fiorino, A., Zhu, L., Thompson, D., *et al.* Nanogap near-field thermophotovoltaics. *Nat Nanotechnol* **13**, 806–811 (2018).

35. Inoue, T., Koyama, T., Kang, D. D., *et al.* One-Chip Near-Field Thermophotovoltaic Device Integrating a Thin-Film Thermal Emitter and Photovoltaic Cell. *Nano Lett* **19**, 3948–3952 (2019).

36. Bhatt, G. R., Zhao, B., Roberts, S., *et al.* Integrated near-field thermo-photovoltaics for heat recycling. *Nat Commun* **11**, 2545 (2020).

37. Lucchesi, C., Cakiroglu, D., Perez, J.-P., *et al.* Near-Field Thermophotovoltaic Conversion with High Electrical Power Density and Cell Efficiency above 14%. *Nano Lett* **21**, 4524–4529 (2021).

38. Yin, X., Yang, R., Tan, G. & Fan, S. Terrestrial radiative cooling: Using the cold universe as a renewable and sustainable energy source. *Science (1979)* **370**, 786–791 (2020).

39. Zhao, B., Hu, M., Ao, X., Chen, N. & Pei, G. Radiative cooling: A review of fundamentals, materials, applications, and prospects. *Appl Energy* **236**, 489–513 (2019).

40. Zhao, B., Hu, M., Ao, X., Chen, N. & Pei, G. Radiative cooling: A review of fundamentals, materials, applications, and prospects. *Appl Energy* **236**, 489–513 (2019).

41. Raman, A. P., Anoma, M. A., Zhu, L., Rephaeli, E. & Fan, S. Passive radiative cooling below ambient air temperature under direct sunlight. *Nature* **515**, 540–544 (2014).

42. Lee, M., Kim, G., Jung, Y., *et al.* Photonic structures in radiative cooling. *Light Sci Appl* **12**, 134 (2023).

43. Kajihara, Y., Kosaka, K. & Komiyama, S. A sensitive near-field microscope for thermal radiation. *Review of Scientific Instruments* **81**, 033706 (2010).

44. Weng, Q., Komiyama, S., Yang, L., *et al.* Imaging of nonlocal hot-electron energy dissipation via shot noise. *Science (1979)* **360**, 775–778 (2018).

45. De Wilde, Y., Formanek, F., Carminati, R., *et al.* Thermal radiation scanning tunnelling microscopy. *Nature* **444**, 740–3 (2006).

46. Jones, A. C. & Raschke, M. B. Thermal infrared near-field spectroscopy. *Nano Lett* **12**, 1475–148 (2012).

47. Huth, F., Schnell, M., Wittborn, J., Ocelic, N. & Hillenbrand, R. Infrared-spectroscopic nanoimaging with a thermal source. *Nat Mater* **10**, 352–356 (2011).

48. Zhang, Z. M. *Nano/Microscale Heat Transfer (2nd Ed.)*. (Springer, 2020).

49. Cuevas, J. C. Thermal radiation from subwavelength objects and the violation of Planck's law. *Nat Commun* **10**, 3342 (2019).

50. Liu, X., Wang, L. & Zhang, Z. M. Near-field thermal radiation: Recent progress and outlook. in *Nanoscale and Microscale Thermophysical Engineering* vol. 19 98–126 (2015).

51. Song, B., Fiorino, A., Meyhofer, E. & Reddy, P. Near-field radiative thermal transport: From theory to experiment. *AIP Adv* **5**, 053503 (2015).

52. Basu, S., Chen, Y.-B. & Zhang, Z. M. Microscale radiation in thermophotovoltaic devices—A review. *Int J Energy Res* **31**, 689–716 (2007).

53. Pascale, M., Giteau, M. & Papadakis, G. T. Perspective on near-field radiative heat transfer. *Appl Phys Lett* **122**, 100501 (2023).

54. Song, J., Han, J., Choi, M. & Lee, B. J. Modeling and experiments of near-field thermophotovoltaic conversion: A review. *Solar Energy Materials and Solar Cells* **238**, 111556 (2022).

55. Overvig, A. C., Mann, S. A. & Alù, A. Thermal Metasurfaces: Complete Emission Control by Combining Local and Nonlocal Light-Matter Interactions. *Phys Rev X* **11**, 21050 (2021).

56. Baranov, D. G., Xiao, Y., Nechepurenko, I. A., *et al.* Nanophotonic engineering of far-field thermal emitters. *Nat Mater* **18**, 920–930 (2019).

57. Ben-Abdallah, P. & Biehs, S. A. Thermotronics: Towards Nanocircuits to Manage Radiative Heat Flux. *Zeitschrift fur Naturforschung - Section A Journal of Physical Sciences* **72**, 151–162 (2017).

58. Biehs, S.-A., Messina, R., Venkataram, P. S., *et al.* Near-field radiative heat transfer in many-body systems. *Rev Mod Phys* **93**, 25009 (2021).

59. Tervo, E., Bagherisereshki, E. & Zhang, Z. Near-field radiative thermoelectric energy converters: a review. *Frontiers in Energy* **12**, 5–21 (2018).

60. Salihoglu, H., Li, Z. & Shen, S. Theory of thermal radiation from a nanoparticle array. *Appl Phys Lett* **121**, 241701 (2022).

61. Sarkar, S., Nefzaoui, E., Hamaoui, G., *et al.* Wideband mid infrared absorber using surface doped black silicon. *Appl Phys Lett* **121**, 231703 (2022).

62. Castillo-López, S. G., Esquivel-Sirvent, R., Villarreal, C. & Pirruccio, G. Near-field radiative heat transfer management by subwavelength plasmonic crystals. *Appl Phys Lett* **121**, 201708 (2022).

63. McSherry, S. & Lenert, A. Design of a gradient epsilon-near-zero refractory metamaterial with temperature-insensitive broadband directional emission. *Appl Phys Lett* **121**, 191702 (2022).

64. Herz, F. & Biehs, S.-A. Thermal radiation and near-field thermal imaging of a plasmonic Su–Schrieffer–Heeger chain. *Appl Phys Lett* **121**, 181701 (2022).

65. Liu, X., Luo, X., Yu, B., *et al.* Electrically driven thermal infrared metasurface with narrowband emission. *Appl Phys Lett* **121**, 131703 (2022).

66. Pouria, R., Chow, P. K., Tiwald, T., Zare, S. & Edalatpour, S. Far-field thermal radiation from short-pitch silicon-carbide nanopillar arrays. *Appl Phys Lett* **121**, 131702 (2022).

67. Liu, Y., Liu, X., Chen, F., *et al.* Intelligent regulation of VO2-PDMS-driven radiative cooling. *Appl Phys Lett* **120**, 171704 (2022).

68. Dumoulin, J., Drouard, E. & Amara, M. Radiative sky cooling of silicon solar modules: Evaluating the broadband effectiveness of photonic structures. *Appl Phys Lett* **121**, 231101 (2022).

69. Kivisaari, P. & Oksanen, J. Resonance effects in the radiation transfer of thin-film intracavity devices. *Appl Phys Lett* **121**, 191101 (2022).

70. Xu, D., Zhao, J. & Liu, L. Electrically tuning near-field heat flux using metal–oxide–semiconductor structure considering gradient dielectric function distribution. *Appl Phys Lett* **121**, 181112 (2022).

71. Pérez-Madrid, A. & Santamaría-Holek, I. Radiative thermal conductance between nanostructures. *Appl Phys Lett* **121**, 181105 (2022).

72. Guo, Y., Zhu, L., Chen, S., *et al.* Dual-band polarized optical switch with opposite thermochromic properties to vanadium dioxide. *Appl Phys Lett* **121**, 201102 (2022).

73. Gharib Ali Barura, C., Ben-Abdallah, P. & Messina, R. Coupling between conduction and near-field radiative heat transfer in tip–plane geometry. *Appl Phys Lett* **121**, 141101 (2022).

74. Tachikawa, S., Ordonez-Miranda, J., Wu, Y., *et al.* In-plane surface phonon-polariton thermal conduction in dielectric multilayer systems. *Appl Phys Lett* **121**, 202202 (2022).

75. Walter, L. P. & Francoeur, M. Orientation effects on near-field radiative heat transfer between complex-shaped dielectric particles. *Appl Phys Lett* **121**, 182206 (2022).

76. Wu, Y., Ordonez-Miranda, J., Jalabert, L., *et al.* Observation of heat transport mediated by the propagation distance of surface phonon-polaritons over hundreds of micrometers. *Appl Phys Lett* **121**, 112203 (2022).

77. Wang, S., Chen, Q. & Hao, Q. Extension of the two-layer model to heat transfer coefficient predictions of nanoporous Si thin films. *Appl Phys Lett* **121**, 012201 (2022).

78. Wang, C.-H., Liu, M.-X. & Jiang, Z.-Y. TiO2 particle agglomeration impacts on radiative cooling films with a thickness of 50 μm. *Appl Phys Lett* **121**, 202204 (2022).

79. Singh, H. K. G., Matsumoto, K. & Sakurai, A. A simple structured solar selective absorber for application in thermoelectric energy harvesters. *Appl Phys Lett* **121**, 173906 (2022).

80. Forcade, G. P., Valdivia, C. E., Molesky, S., *et al.* Efficiency-optimized near-field thermophotovoltaics using InAs and InAsSbP. *Appl Phys Lett* **121**, 193903 (2022).

81. Legendre, J. & Chapuis, P.-O. Overcoming non-radiative losses with AlGaAs PIN junctions for near-field thermophotonic energy harvesting. *Appl Phys Lett* **121**, 193902 (2022).

82. Chirtoc, M. & Horny, N. Correlation between amplitude and phase photothermal signals based on Kramers–Kronig relations: Application to the investigation of interfacial thermal resistance in multilayers. *Appl Phys Lett* **121**, 183504 (2022).

83. Song, J., Choi, M., Yang, Z., Lee, J. & Lee, B. J. A multi-junction-based near-field solar thermophotovoltaic system with a graphite intermediate structure. *Appl Phys Lett* **121**, 163503 (2022).

84. Liu, Z., Abe, S., Shimizu, M. & Yugami, H. Enhanced current density and asymmetry of metal–insulator–metal diodes based on self-assembly of Pt nanoparticles. *Appl Phys Lett* **122**, 093502 (2023).

85. Ning, D., Xu, L., Zhu, Y., *et al.* Bessel beam induced deep-penetrating bioimaging and self-monitored heating using Nd/Yb heavily doped nanocrystals. *Appl Phys Lett* **121**, 043701 (2022).

86. Zhang, Y. & Fang, H. Surface-enhanced photoacoustic wave generation from light absorbers located in the gap of high-refractive-index dielectric nanoparticles. *Appl Phys Lett* **121**, 213701 (2022).

87. Luk, T. S., Xu, G., Ross, W., *et al.* Maximal absorption in ultrathin TiN films for microbolometer applications. *Appl Phys Lett* **121**, 234101 (2022).

88. Dedkov, G. V. Nonequilibrium Casimir–Lifshitz friction force and anomalous radiation heating of a small particle. *Appl Phys Lett* **121**, 231603 (2022).